\documentclass[superscripaddress, floatfix, showkeys, nofootinbib]{revtex4-2}
\usepackage{times,fancyhdr}
\usepackage[dvips]{graphicx}
\usepackage{amsmath,amssymb,bm}
\usepackage{color}
\usepackage{mathrsfs}
\usepackage{setspace}
\usepackage{hyperref}
\usepackage{natbib}
\usepackage{amsfonts}
\usepackage{subfig}
\usepackage{amsmath,amssymb,graphicx,geometry,physics}
\usepackage{circuitikz}
\begin{document}
\title{Dimensional consistency in fractional differential equations with non singular kernels}
\author{Gabriel Gonz\'alez}
\affiliation{Investigador por M\'exico-Departamento de Ciencias B\'asicas y Aplicadas, CUTonalá, Universidad de Guadalajara Av. Nuevo Perif\'erico No. 555 Ejido San José Tateposco, Tonal\'a 45425, M\'exico}
\email{gabriel.gonzalez@academicos.udg.mx}
\begin{abstract}
{\bf Abstract}\\
  The purpose of this article is to address the issues of dimensional consistency that arise in the process of replacing the ordinary time derivative operator by a fractional derivative operator in order to write a fractional differential equation. We show that by performing a simple change of variables fulfilling certain conditions ensures the consistency in physical dimensions for fractional differential equations with non singular kernels. An example of the proposed method is given. 
\end{abstract}
\vspace{-8pt}

\maketitle
\section{Introduction}
From a mathematical point of view, fractional calculus was introduced as a generalization of classical calculus and has become a powerful tool for modeling complex dissipative phenomena, attracting the interest of scientists and being currently the subject of many theoretical and experimental investigations \cite{kochubei2019handbook, amilo2025integrated, iqbal2024new}. Among the various definitions of fractional derivatives, Riemann–Liouville and Caputo operators, based on power-law kernels, are widely used to capture memory effects with algebraic decay \cite{yavuz2020comparing}. However, these kernels are singular at the origin, which can lead to difficulties in handling initial conditions and interpreting physical processes. To overcome these limitations, Caputo and Fabrizio proposed a fractional derivative with smooth non-singular exponential kernels, which have the ability to describe material heterogeneities and multi-scale fluctuations\cite{caputo2015new}. Despite these kernel design development, incorporating fractional derivatives into physical models introduces additional challenges related to dimensional consistency. \\
%The modelling of physical systems with dissipative forces, mass and heat transfer behaviour in many new emerging materials was more accurate using non singular kernels in fractional derivatives in contrast to the ones using fractional derivatives with singular kernels.\\
Fractional derivatives, whether defined with singular or non-singular kernels, are incorporated into differential equations to model physical phenomena through fractional differential equations (FDEs). Through these formulations, non-integer models have been applied to systems such as the parallel RLC circuit or the damped harmonic oscillator \cite{da2020fractional}. However, the straightforward replacement of an integer order derivative by a fractional order derivative in equations involving physical variables leads to dimensional imbalance that raises questions about the interpretability of the models \cite{vaz2025fractionaldifferentialequationsdimensional}.\\To address this issue, some strategies are implemented, such as introducing a scaling parameter or characteristic constant that restores the dimensional homogeneity in the equations. For example, Gómez-Aguilar et al. \cite{gomez2012fractional,correa2022correcting}, and Bachuin\cite{Banchuin_2019} introduced an auxiliary parameter $\sigma$ that represents the fractional time components of the system, naming the non-local time as the \textit{cosmic time}. This parameter allows the fractional derivative to be dimensionally consistent with its classical counterpart, even though its physical interpretation remains challenging, as it typically carries units such as sec$^{-\alpha}$ that are not directly measurable in practice.\\
Alternatively, Vaz and Capelas de Oliveira \cite{vaz2025fractionaldifferentialequationsdimensional} describe a dimensional regularization technique for the fractional derivative of Caputo, where a multiplicative term involving the independent variable is introduced to maintain the original physical dimension of the problem. This approach avoids the arbitrariness of auxiliary parameters by systematically balancing the dimensions through characteristic scales. \\
Other methods include normalization procedures that transform variables into dimensionless forms to avoid dimensional balancing, but when reintroducing physical dimensions into fractional form, some problems are encountered when interpreting the terms of the FDE \cite{vaz2025fractionaldifferentialequationsdimensional}.\\
The main goal of this article is to answer the question on how physical dimensions in a fractional derivative operator has to be taken into account to model real problems and to have dimensional consistency in the fractional differential system.

\section{Dimensional consistency in fractional derivative operators}
The simplest procedure for constructing a fractional differential equation consists of simply replacing all conventional derivative operators of a given system by the fractional derivative operator. The replacement of an integer order derivative operator by a fractional derivative operator in a differential equation leads to a dimensional inconsistency. For example, let us consider the definition of the classical Caputo fractional derivative, which is given by
\begin{equation}\label{eq1}
  {}^{C}D_t^{\alpha}f(t)=\frac{1}{\Gamma(1-\alpha)}\int_{a}^{1}\frac{df(s)}{ds}\frac{1}{(t-s)^{\alpha}}ds
\end{equation}
with $\alpha\in [0,1]$ and $a\in [-\infty,t]$ and $f(t)\in H^{1}(a,b)$ for $b>a$. 
If we do such replacements we have a dimensional inconsistency since the time dimension of ${}^{C}D_t^{\alpha}$, i.e. the Caputo fractional derivative, is sec$^{-\alpha}$ while that of $d/dt$ is sec$^{-1}$. For achieving dimensional consistency, G\'omez \textit{et al} . introduce the following simple rule which consists of replacing the ordinary time derivative operator by the fractional one in the following way
\begin{align}\label{eq01}
  \frac{d}{dt}\rightarrow & \frac{1}{\sigma^{1-\alpha}}  {}^{C}D_t^{\alpha} \quad \mbox{where $0<\alpha\leq1$} 
\end{align}
where $\sigma>0$ is a new parameter which has the dimension of time, i.e. $[\sigma]=s$. Using this procedure, it is easy to write down the fractional differential equation of any given system. Note that now the dimensions of $d/dt$ and $(1/\sigma^{1-\alpha})\,{}^{C}D_t^{\alpha}$ are now consistent.\\
However, there are many different definitions of fractional derivatives in the literature. Traditional definitions of fractional derivatives, such as the Caputo operator, employ power-law kernels with singularities at the origin. Although these kernels effectively represent long-range memory, their singular behavior often leads to analytical difficulties, numerical instabilities, and ambiguity in the physical interpretation of initial conditions. 
In order to overcome these limitations, Caputo and Fabrizio proposed a new fractional derivative in 2015 characterized by a non-singular and nonlocal exponential kernel. This formulation preserves the memory effect inherent to fractional calculus while ensuring improved regularity and computational tractability. Caputo and Fabrizio proposed changing the kernel $(t-s)^{-\alpha}$ in the Caputo fractional derivative by the exponential function $\exp[-(\alpha t/(1-\alpha))]$ and $1/\Gamma(1-\alpha)$ by $(2-\alpha)M(\alpha)/2(1-\alpha)$, to obtain the following definition for the fractional derivative with non singular kernel of the Caputo-Fabrizio type given by
\begin{equation}\label{eq2}
{}^{CF}D_t^{\alpha}f(t)=\frac{(2-\alpha)M(\alpha)}{2(1-\alpha)}\int_{0}^{t}\frac{df(s)}{ds}\exp\left[-\frac{\alpha(t-s)}{1-\alpha}\right]ds  
\end{equation} 
where $M(\alpha)$ is a normalization function such that $M(\alpha)=2/(2-\alpha)$. \\
The Caputo--Fabrizio derivative satisfies several fundamental properties that make it suitable for theoretical and applied mathematical physics. The Caputo--Fabrizio operator satisfies the following mathematical properties

\paragraph{Linearity.}
For any functions $f,g \in C^{1}([0,T])$ and constants $a,b \in \mathbb{R}$,
\begin{equation}
{}^{CF}D_t^{\alpha} \left( a f(t) + b g(t) \right)
=
a\, {}^{CF}D_t^{\alpha} f(t)
+
b\, {}^{CF}D_t^{\alpha} g(t).
\end{equation}

\paragraph{Derivative of a Constant.}
If $f(t)=C$, where $C$ is a constant, then
\begin{equation}
{}^{CF}D_t^{\alpha} C = 0,
\end{equation}
which is consistent with classical differentiation and advantageous for physical modeling.

\paragraph{Limit Cases.}
The Caputo--Fabrizio derivative interpolates between classical operators:
\begin{equation}
\lim_{\alpha \to 0} {}^{CF}D_t^{\alpha} f(t) = f(t) - f(0),
\qquad
\lim_{\alpha \to 1} {}^{CF}D_t^{\alpha} f(t) = \frac{d f(t)}{dt}.
\end{equation}

\paragraph{Laplace Transform.}
Let $\mathcal{L}\{f(t)\}(s)=F(s)$. The Laplace transform of the Caputo--Fabrizio derivative is given by
\begin{equation}
\mathcal{L}\left\{ {}^{CF}D_t^{\alpha} f(t) \right\}(s)
=
\frac{1}{1-\alpha}
\frac{s F(s) - f(0)}{s + \frac{\alpha}{1-\alpha}},
\end{equation}
which facilitates the analytical solution of fractional differential equations.\\
Due to its non-singular kernel, well-defined initial conditions, and favorable analytical properties, the Caputo--Fabrizio derivative has been extensively employed in the formulation and analysis of fractional differential equations. Applications include stability analysis of dynamical systems, reaction--diffusion processes, and control problems, where it provides a mathematically consistent and physically interpretable alternative to classical fractional operators. The operator \eqref{eq2} is nonlocal, as the value of the derivative at time $t$ depends on the entire history of the function over $[0,t]$. However, unlike power-law kernels, the exponential kernel induces a memory with finite persistence, which is often more realistic in applied models.\\
From equation (\ref{eq2}) we see that the Caputo-Fabrizio fractional derivative is a dimensionless operator, therefore in order to write down the corresponding fractional differential operator we need to introduce a new dimensionless parameter given by
\begin{equation}\label{eq02}
  \tau(t,\alpha)=\int_0^t\frac{dt}{\varphi(t,\alpha)}
\end{equation}
where $\varphi(t,\alpha)$ is a function that has the dimension of time and depends on the order of the fractional derivative. Therefore, we can replace the ordinary time derivative operator by the following one
\begin{equation}\label{eq03}
  \frac{d}{dt}\rightarrow \frac{d\tau}{dt}\frac{d}{d\tau}\rightarrow \frac{1}{\varphi(t(\tau),\alpha)} {}_{CF}D_t^{\alpha}
\end{equation} 
In order to obtain the ordinary time derivative in equation (\ref{eq03}) when $\alpha=1$ we must impose the following condition: 
\begin{equation}\label{eq04}
  \frac{1}{\varphi(t,1)}\frac{d}{d\tau}=\frac{d}{dt}
\end{equation}
Therefore, if we would like to solve the fractional differential equation, using the Caputo-Fabrizio fractional derivative operator, equivalent to a first order differential equation with constant coefficients given by
\begin{equation}\label{eq5}
    \frac{dx}{dt}+\mathrm{P} x(t)=\mathrm{Q}
\end{equation}
where $\mathrm{P}$ and $\mathrm{Q}$ are constants, we proceed to solve the following fractional differential equation
\begin{equation}\label{eq6}
    {}^{CF}D_{\tau}^{\alpha}x(\tau)+\mathrm{P}(\tau)x(\tau)=\mathrm{Q}(\tau)
\end{equation}
where $\mathrm{P}(\tau)=\mathrm{P}\varphi(t(\tau),\alpha)$ and $\mathrm{Q}(\tau)=\mathrm{Q}\varphi(t(\tau),\alpha)$. Substituting equation (\ref{eq2}) into equation (\ref{eq6}) we have
\begin{equation}\label{eq7}
    \frac{e^{-\alpha\tau/(1-\alpha)}}{1-\alpha}\int^{\tau}_{0}e^{\alpha u/(1-\alpha)}x^{\prime}(u)du+\mathrm{P}(\tau)x(\tau)=\mathrm{Q}(\tau)
\end{equation}
Differentiating both sides of equation (\ref{eq7}) with respect to $\tau$ and simplifying, we get
\begin{equation}\label{eq8}
    \frac{dx}{d\tau}=(1-\alpha)\frac{d}{d\tau}\left(\mathrm{Q}(\tau)-\mathrm{P}(\tau)x(\tau)\right)+\alpha\left(\mathrm{Q}(\tau)-\mathrm{P}(\tau)x(\tau)\right)
\end{equation}
which can be rearranged as
\begin{equation}\label{eq9}
    \left((1-\alpha)\mathrm{P}(\tau)+1\right)x^{\prime}(\tau)+\left((1-\alpha)\mathrm{P}^{\prime}(\tau)+\alpha \mathrm{P}(\tau)\right)x(\tau)-\left((1-\alpha)\mathrm{Q}^{\prime}(\tau)+\alpha \mathrm{Q}(\tau)\right)=0
\end{equation}
Equation (\ref{eq9}) has the following integrating factor \cite{0}
\begin{equation}\label{eq9a}
    \mu(\tau)=\exp{\int\frac{\alpha \mathrm{P}(\tau)}{1+(1-\alpha)\mathrm{P}(\tau)}d\tau}
\end{equation}
Therefore, multiplying equation (\ref{eq9}) by $\mu(\tau)$ and integrating we arrive to the solution of equation (\ref{eq6}) which is given by
\begin{equation}\label{eq9b}
    x(\tau)=\Xi(\tau)\left(C+\int\mu(\tau)\left((1-\alpha)\mathrm{Q}^{\prime}(\tau)+\alpha \mathrm{Q}(\tau)\right)\right)
\end{equation}
where
\begin{equation}\label{eq9c}
    \Xi(\tau)=\frac{1}{\left((1-\alpha)\mathrm{P}(\tau)+1\right)\mu(\tau)}
\end{equation}
\section{Example: Analysis of an RC circuit}
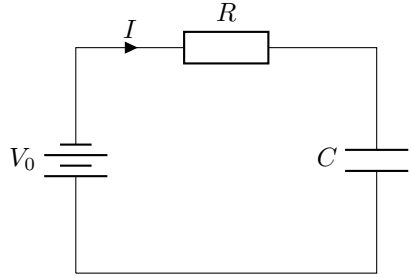
\begin{figure}[h]
\centering
%\begin{circuitikz}
%\node[spdt, rotate=90] (sw) {};
%\draw   (sw.in)     to [R=$R$, european] ++ (0,-3)
%                    to [C,l_=$C$] ++ (0,-3) coordinate (aux1)
%        (sw.out 2)  node[above] {(2)}   to [short] ++ (+2,0) %|- (aux1)
%        (sw.out 1)  node[above] {(1)}   to [short] ++ (-2,0) %coordinate (aux2)
%                    to [battery, a=$V_0$]    (aux2 |- aux1)
%                    to [short] (aux1);
%    \end{circuitikz}
\begin{circuitikz}
\draw
(0,0)
to [battery, l=$V_0$] (0,3)
to [R=$R$, european,i>^=$I$] (4,3)
to [C,l_=$C$] (4,0)
-- (0,0);
\end{circuitikz}
\caption{RC circuit.}
\label{fig:RCcircuit}
\end{figure}

We are now interested in the study of a simple electrical circuit consisting of a resistor and a capacitor within the framework of the fractional derivative with a non singular kernel \cite{1,2,3,gomez2014physical,gomez2015modeling,ertik2015investigation}. 
Using the Kirchhoff voltage law in the circuit shown in Fig.~\ref{fig:RCcircuit}, we get the following fractional differential equation  

\begin{equation}\label{eq05}
 {}^{CF}D_{\tau}^{\alpha}q(t)+\Gamma q(t)=\Gamma q_0
\end{equation}
where $\Gamma=1/RC$ and $q_0=V_0C$. \\

Several authors include an auxiliary parameter $\sigma$ which has the dimension of seconds and is associated with the temporal components in the system in order to keep the dimensional consistency of the fractional differential equation, which is valid for the case of singular kernels \cite{4,5,6,7,8,9,10}. \\
For the case of differential equations with non singular kernels we have to introduce an auxiliary function $\varphi$ in order to have dimensional consistency in equation (\ref{eq05}), therefore

\begin{equation}\label{eq06}
 {}^{CF}D_{\tau}^{\alpha}q(\tau)+\frac{q}{1+(1-\alpha)\tau}=\frac{q_0}{1+(1-\alpha)\tau}\quad \mbox{where $0<\alpha\leq 1$}
\end{equation}
where we have taken $\varphi(t,\alpha)=e^{-(1-\alpha)\Gamma t}/\Gamma$ and using equation (\ref{eq02}) we have
\begin{equation}\label{eq07}
  \tau(t,\alpha)=\frac{e^{(1-\alpha)\Gamma t}-1}{1-\alpha} 
\end{equation}
Note that equation (\ref{eq04}) is satisfied. 
Substituting $\mathrm{P}(\tau)=[1+(1-\alpha)\tau]^{-1}$ into equation (\ref{eq9a}) and using equations (\ref{eq9b}) and (\ref{eq9c}) with $\mathrm{Q}(\tau)=q_0[1+(1-\alpha)\tau]^{-1}$ we find the solution of the fractional differential equation given in equation (\ref{eq06}) which is given by
\begin{equation}\label{eq08}
  q(\tau,\alpha)=q_0+C_0(1+(1-\alpha)\tau)\left(2-\alpha+(1-\alpha)\tau\right)^{1/(\alpha-1)}
\end{equation}
Using the initial condition $q(0,\alpha)=0=q_0+C_0(2-\alpha)^{1/(\alpha-1)}$ and
using equation (\ref{eq07}) we can write the solution in terms of the ordinary time and the initial condition which is given by
\begin{equation}\label{eq09}
  q(t,\alpha)=q_0\left(1-e^{(1-\alpha)\Gamma t}\left(\frac{2-\alpha}{(1-\alpha)+e^{(1-\alpha)\Gamma t}}\right)^{1/(1-\alpha)}\right)
\end{equation}
It is easy to verify the solution by taking $\alpha\rightarrow1$ in equation (\ref{eq09}) in order to get
\begin{equation}\label{eq10}
  q(t,1)=q_0\left(1-e^{-\Gamma t}\right)
\end{equation}
which is the expected solution for an RC circuit connected to a DC battery. \\
The voltage across the capacitor while it is charging is given by
\begin{equation}\label{eq10}
    V_C(t,\alpha)=V_0\left(1-e^{(1-\alpha)\Gamma t}\left(\frac{2-\alpha}{(1-\alpha)+e^{(1-\alpha)\Gamma t}}\right)^{1/(1-\alpha)}\right)
\end{equation}

\begin{figure}
      \centering
      \includegraphics[width=0.70\linewidth]{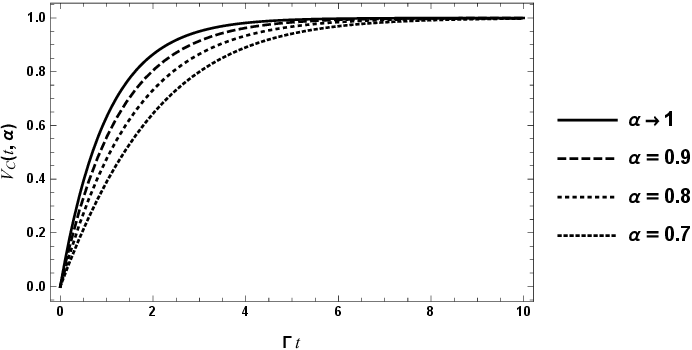}
      \caption{The figure shows the voltage across the capacitor }\label{fig01}
      \label{fig:placeholder}
  \end{figure}
In figure (\ref{fig01}) we show the graph of the voltage across the capacitor given by equation (\ref{eq10}) for different values of $\alpha$.
The results plot given in figure (\ref{fig01}) shows that when $\alpha\rightarrow1$ the system displays the classical behavior. However, for values of $0<\alpha<1$, the solution describes a dissipative system due to the fact that the fractional differentiation with respect
to the time represents a non-local effect of dissipation of energy (internal friction) represented by the fractional order $\alpha$.  We should point out that there might be other functions $\varphi$ that could result in a different fractional differential equation and that will result in other classes of solutions. 
\vspace{-1cm}
\section{Conclusions}
A systematic way to construct fractional differential equations with non singular kernels has been proposed. The relevant aspect of this work is the method for setting up the fractional differential equations with nonsingular kernels while keeping the dimensionality of the physical parameters. A new function has been introduced, that depends on the order of the derivative being considered i.e. $0<\alpha\leq1$, to keep the dimensionality of the physical parameters. An example of this new approach has been given for the case of an RC circuit connected to a DC battery. The results presented in this paper can be extended to non singular fractional derivatives of higher order.

\section{Acknowledgement}
I would like to acknowledge support from the program IXM through project 6999. I will also like to thank useful discussions with Dr. Guillermo Fern\'andez and Dra. Milagros C. Santos at the beginning of this work.  
\bibliographystyle{unsrt}
\bibliography{ref.bib}

@misc{vaz2025fractionaldifferentialequationsdimensional,
      title={On fractional differential equations, dimensional analysis, and the double gamma function},
      author={J. Vaz and E. Capelas de Oliveira},
      year={2025},
      eprint={2506.00026},
      archivePrefix={arXiv},
      primaryClass={math.GM},
      url={https://arxiv.org/abs/2506.00026},
}

@ARTICLE{Banchuin_2019,
title={Time Dimensional Consistency Aware Analysis of Voltage Mode and Current Mode Active Fractional Circuits},
year={2019},author={Rawid Banchuin and Rawid Banchuin and Roungsan Chaisrichaoren and Roungsan Chaisrichaoren},
doi={10.37936/ecti-cit.2019131.123094},
journal={ECTI Transactions on Computer and Information Technology}}

@article{da2020fractional,
  title={Fractional derivative order determination from harmonic oscillator damping factor},
  author={da Silva, Lu{\'\i}s Felipe Alves and J{\'u}nior, Valdiney Rodrigues Pedrozo and Ferreira, Jo{\~a}o V{\'\i}tor Batista},
  journal={Chinese Journal of Physics},
  volume={66},
  pages={673--683},
  year={2020},
  publisher={Elsevier}
}

@article{gomez2012fractional,
  title={Fractional mechanical oscillators},
  author={G{\'o}mez-Aguilar, JF and Rosales-Garc{\'\i}a, JJ and Bernal-Alvarado, JJ and C{\'o}rdova-Fraga, T and Guzm{\'a}n-Cabrera, R},
  journal={Revista mexicana de f{\'\i}sica},
  volume={58},
  number={4},
  pages={348--352},
  year={2012},
  publisher={Sociedad Mexicana de F{\'\i}sica}
}

@article{correa2022correcting,
  title={Correcting dimensional mismatch in fractional models with power, exponential and proportional kernel: Application to electrical systems},
  author={Correa-Escudero, IL and G{\'o}mez-Aguilar, JF and L{\'o}pez-L{\'o}pez, MG and Alvarado-Mart{\'\i}nez, VM and Baleanu, D},
  journal={Results in Physics},
  volume={40},
  pages={105867},
  year={2022},
  publisher={Elsevier}
}

@book{kochubei2019handbook,
  title={Handbook of fractional calculus with applications},
  author={Kochubei, Anatoly and Luchko, Yuri and Tarasov, Vasily E and Petr{\'a}{\v{s}}, Ivo},
  volume={1},
  year={2019},
  publisher={de Gruyter Berlin}
}

@article{amilo2025integrated,
  title={An integrated machine learning and fractional calculus approach to predicting diabetes risk in women},
  author={Amilo, David and Sadri, Khadijeh and Hincal, Evren and Farman, Muhammad and Nisar, Kottakkaran Sooppy and Hafez, Mohamed},
  journal={Healthcare Analytics},
  pages={100402},
  year={2025},
  publisher={Elsevier}
}

@article{iqbal2024new,
  title={New applications of the fractional derivative to extract abundant soliton solutions of the fractional order PDEs in mathematics physics},
  author={Iqbal, M Ashik and Miah, M Mamun and Ali, HM Shahadat and Shahen, Nur Hasan Mahmud and Deifalla, Ahmed},
  journal={Partial Differential Equations in Applied Mathematics},
  volume={9},
  pages={100597},
  year={2024},
  publisher={Elsevier}
}

@article{caputo2015new,
  title={A new definition of fractional derivative without singular kernel},
  author={Caputo, Michele and Fabrizio, Mauro},
  journal={Progress in fractional differentiation \& applications},
  volume={1},
  number={2},
  pages={73--85},
  year={2015}
}

@article{yavuz2020comparing,
  title={Comparing the new fractional derivative operators involving exponential and Mittag-Leffler kernel},
  author={Yavuz, Mehmet and {\"O}zdemir, Necati},
  journal={Discret. Contin. Dyn. Syst.-S},
  volume={13},
  pages={995},
  year={2020}
}

@article{0,
  title={On solutions of linear and nonlinear fractional differential equations with application to fractional order RC type circuits},
  author={AlAhmad, Rami and Al-Khaleel, Mohammad and Almefleh, Hasan},
  journal={Journal of Computational and Applied Mathematics},
  volume={438},
  pages={115507},
  year={2024},
  publisher={Elsevier}
}

@article{1,
  title={Analyzing transient response of the parallel RCL circuit by using the Caputo–Fabrizio fractional derivative},
  author={Alizadeh, Shahram and Baleanu, Dumitru and Rezapour, Shahram},
  journal={Advances in Difference Equations},
  volume={2020},
  pages={55},
  year={2020},
  doi={10.1186/s13662-020-2527-0},
}

@article{2,
  title={Electrical circuits described by a fractional derivative with regular kernel (Caputo–Fabrizio)},
  author={Gómez-Aguilar, J.F. and Córdova-Fraga, T. and Escalante-Martínez, J.E. and others},
  journal={Revista Mexicana de Física},
  volume={62},
  number={2},
  pages={144--154},
  year={2016},
}

@article{3,
  title={A comparative mathematical analysis of RL and RC electrical circuits via Atangana–Baleanu and Caputo–Fabrizio fractional derivatives},
  author={Abro, K.A. and Memon, A.A. and Uqaili, M.A.},
  journal={European Physical Journal Plus},
  volume={133},
  pages={113},
  year={2018},
}

@article{4,
  title={A comparative analysis of the RC circuit with local and non-local fractional derivatives},
  author={Rosales García, J.J. and Filoteo, J.D. and González, A.},
  journal={Revista Mexicana de Física},
  volume={64},
  number={6},
  pages={647--654},
  year={2018},
  doi={10.31349/RevMexFis.64.647},
}

@article{ertik2015investigation,
  title={Investigation of electrical RC circuit within the framework of fractional calculus},
  author={Ertik, H{\"U}SEY{\.I}N and Calik, AE and {\c{S}}irin, HAT{\.I}CE and {\c{S}}en, M and {\"O}der, B},
  journal={Revista mexicana de f{\'\i}sica},
  volume={61},
  number={1},
  pages={58--63},
  year={2015},
  publisher={Sociedad Mexicana de F{\'\i}sica}
}

@article{gomez2014physical,
  title={A physical interpretation of fractional calculus in observables terms: analysis of the fractional time constant and the transitory response},
  author={G{\'o}mez-Aguilar, Jos{\'e} Francisco and Razo-Hern{\'a}ndez, R and Granados-Lieberman, D},
  journal={Revista mexicana de f{\'\i}sica},
  volume={60},
  number={1},
  pages={32--38},
  year={2014},
  publisher={Sociedad Mexicana de F{\'\i}sica}
}

@article{gomez2015modeling,
  title={Modeling of a mass-spring-damper system by fractional derivatives with and without a singular kernel},
  author={G{\'o}mez-Aguilar, Jos{\'e} Francisco and Y{\'e}pez-Mart{\'\i}nez, Huitzilin and Calder{\'o}n-Ram{\'o}n, Celia and Cruz-Ordu{\~n}a, Ines and Escobar-Jim{\'e}nez, Ricardo Fabricio and Olivares-Peregrino, Victor Hugo},
  journal={Entropy},
  volume={17},
  number={9},
  pages={6289--6303},
  year={2015},
  publisher={MDPI}
}

@article{5,
  title={Electrical circuits RC, LC, and RL under generalized fractional derivatives},
  author={Acay, B. and Bas, E. and Abdeljawad, T.},
  journal={Eur. Phys. J. Plus},
  volume={136},
  pages={1--14},
  year={2021},
}

@article{6,
  title={Fundamental solutions to electrical circuits of non-integer order via fractional time derivatives},
  author={Gómez-Aguilar, J.F. and Yépez-Martínez, H. and Escobar-Jiménez, R.F. and others},
  journal={European Physical Journal Plus},
  volume={133},
  pages={197},
  year={2018},
}

@article{7,
  title={Analytical solutions of the electrical RLC circuit via Liouville–Caputo and regular kernels},
  author={Morales-Delgado, V.F. and Taneco-Hernández, M.A. and Al Qurashi, M. and Gómez-Aguilar, J.F.},
  journal={Entropy},
  volume={18},
  pages={402},
  year={2016},
}

@article{8,
  title={Analog Realization of Fractional-Order Capacitor and Inductor via the Caputo–Fabrizio Derivative},
  author={Ran, M. and Liao, X. and Lin, D. and Yang, R.},
  journal={J. Appl. Comput. Intell.},
  volume={25},
  number={3},
  pages={291--300},
  year={2021},
  doi={10.20965/jaciii.2021.p0291},
}

@article{9,
  title={Fractional input stability for electrical circuits described by the RC and RL equations},
  author={Sene, N.},
  journal={AIMS Mathematics},
  volume={4},
  number={1},
  pages={147--165},
  year={2019},
}

@article{10,
  title={Fractional-order RC charging circuit experiments based on Caputo–Fabrizio and Atangana–Baleanu derivatives},
  author={Liao, X. and others},
  journal={Fractals},
  volume={29},
  pages={2150235},
  year={2021},
}

\end{document}